\documentclass[reqno]{article}
\usepackage{amssymb,amsmath,epsfig,cite}

\begin{document}
\begin{center}{\Large \bf A five-dimensional perspective on the Klein-Gordon equation}
\vskip.25in
Romulus Breban

{\it Institut Pasteur, 75724 Paris Cedex 15, France} %
\end{center}

\begin{abstract} 
We discuss the Klein-Gordon (KG) equation using a 5D space-time approach. We explicitly show that the KG equation in flat space-time admits a consistent probabilistic interpretation with positively defined probability density.   However, the probabilistic interpretation is not covariant.  In the non-relativistic limit, the formalism reduces naturally to that of the Schr\"odinger equation.  We further discuss other interpretations of the KG equation (and their non-relativistic limits) resulting from the 5D space-time picture.  Finally, we apply our results to the problem of hydrogenic spectra and calculate the canonical sum of  the hydrogenic atom.
\end{abstract}



\section{Introduction}
\label{sec:intro}

The Klein-Gordon (KG) equation describes the quantum propagation of a spineless particle in a flat four-dimensional (4D) space-time.  We denote the flat 4D metric by $\eta_{\mu\nu}\equiv\mbox{\it diag}(-1,1,1,1)$ where $\mu,\nu=0,1,2,3$ and introduce cartesian coordinates $x^\mu$, where $x^0$ may also be written as $x^0=ct$ ($c$ is the speed of light and $t$ is called {\it coordinate time}). With the notation $\partial_\mu$ for the partial derivative with respect to $x^\mu$, the KG equation for a free particle with mass $m$ is written as
\begin{eqnarray}
\partial^\mu\partial_\mu\Psi-\frac{m^2c^2}{\hbar^2}\Psi=0.
\label{eq:KG}
\end{eqnarray}
A conserved four-current $j_\mu$ is defined as ($*$ is symbol for complex conjugation) 
\begin{eqnarray}
j_\mu \propto (\Psi^*\partial_\mu\Psi-\Psi\partial_\mu\Psi^*).
\label{eq:j}
\end{eqnarray}
A local U(1) transformation of $\Psi$, equivalent to a gauge transformation of the electromagnetic field, indicates how electromagnetic interactions enter the KG equation, according to the {\it minimal coupling} recipe.  In the Lorentz gauge, the KG equation with electromagnetic field becomes
\begin{eqnarray}
\left(\partial^\mu-i\frac{e}{c\hbar}A^\mu\right)\left(\partial_\mu-i\frac{e}{c\hbar}A_\mu\right)\Psi-\frac{m^2c^2}{\hbar^2}\Psi=0,
\label{eq:KG-EM}
\end{eqnarray}
where $A_\mu$ is the electromagnetic vector potential, $e$ is the electromagnetic charge of the particle, and $i$ is the complex unit.\footnote{In the presence of the electromagnetic field, we obtain $\partial^\mu j_\mu\propto\partial^\mu(A_\mu\Psi^*\Psi)$. Hence, some authors \cite{Gross:2008vj, Greiner:2008wv} change the definition of $j_\mu$ by a term proportional to $A_\mu\Psi^*\Psi$ so that newly defined current density has zero divergence.  However, it is unclear whether a fix is needed.  As is, current conservation already holds for scattering systems and stationary states (i.e., $\Psi^*\Psi$ is constant at infinity), and situations where the electromagnetic field vanished at infinity.}

The probabilistic, quantum-mechanical interpretation of the KG equation met with difficulties; e.g., Refs.~\cite{Schweber:2009by,Blokhintsev:2009uk}. For this reason, the KG equation was regarded with skepticism in the early days of quantum mechanics \cite{KRAGH:1984we}. Subsequent work, providing an alternate interpretation \cite{Pauli:1934dy} and showing that the Schr\"odinger equation can be obtained as the non-relativistic limit of the KG equation \cite{1958RvMP...30...24F}, revived interest in the KG equation.  There are two major approaches to taking the non-relativistic limit of the KG equation.

One approach proceeds by separating the {\it small} and {\it large} components of the wavefunction; then particle and antiparticle states, satisfying the Schr\"odinger equation, are identified using the Foldy-Wouthysen procedure \cite{1958RvMP...30...24F, Bjorken:1964yg, Gross:2008vj, Grandy:1990wy, Greiner:2000tw}. In the other approach, one solves for the partial derivative of the wavefunction with respect to time using a square root, which is further expanded in Taylor series. The Sch\"odinger equation results in the limit $c\rightarrow\infty$ \cite{Yndurain:2011vu, Strange:1998vn, Moss:1973ta}. Claims are that the two approaches are equivalent \cite{Wu:1991ta}.  More recent work \cite{Bechouche:2004wl,Masmoudi:2003ve} shows that, in the limit $c\rightarrow\infty$, the system formed by the KG and Maxwell equations (where $j_\mu$ is interpreted as electric current density and serves as coupling) yields, as leading order, the Schr\"odinger equation together with the Poisson equation for the electric potential, $A_0$.  A third approach, less popular, is performing the asymptotic expansion when the kinetic energy is much smaller than the rest energy, $E-mc^2\ll mc^2$ \cite{Greiner:2000tw}.  A discussion of the KG equation using initial and final boundary conditions for the scalar field, and touching ground with the Schr\"odinger equation, can be found in Ref.~\cite{2010FoPh...40..313W} and references therein.
 
Here we re-evaluate the KG equation using a 5D space-time with a space-like fifth dimension that is neither compact nor Planckian.  Such geometries have been previously studied \cite{Wesson:2007ta}.  Following Kaluza's idea \cite{1921SPAW.......966K}, it has been shown that, if the 5D geometry is independent of the fifth coordinate, the 5D gravitational equations break into 4D equations for the gravitational and electromagnetic fields \cite{Wesson:2007ta}.  Furthermore a discussion of geodesic motion and classical tests of 4D general relativity adapted to this 5D space-time geometry are summarized in Ref.~\cite{Wesson:2007ta}.  However, our interpretation of the 5D geometry \cite{Breban:2005a} is different from that of other authors. We assume that an observer perceives 4D geometrically, which, in principle, could be any four dimensions out of five, while the extra dimension manifests indirectly. Hence, in our case, the relation between our theory and experiment does not follow the paved way of the classical tests of general relativity.
   
 In Ref.~\cite{Breban:2005a}, we used a 5D space-time to show both how the KG equation emerges from a relativistic path integral (Sec.~5), and the Schr\"odinger equation emerges from the non-relativistic limit of the same path integral (Sec.~3.2).  Here we demonstrate explicitly that the Schr\"odinger equation results by taking the non-relativistic limit of the KG equation. We reinterpret the KG equation, revisiting its probabilistic, quantum-mechanical interpretation.

\section{The KG equation in 5D space-time}
\label{sec:5DKG}

We postulate that, given a space-time with the metric $h_{AB}=\mbox{\it diag}(-1,1,1,1,1)$ where $A,B, ... =0,1,2,3,5$  ($\mu,\nu, ...$ go over 4 out of the 5 values $0,1,2,3,5$---to be specified---and $j,k, ... = 1,2,3$), all quantum propagation takes place on 5D null paths.  Consider any two causally ordered events 1 and 2, with 1 in the past of 2, which we write as $1\prec 2$. Denote the coordinates of 1 and 2 by $x^A_{(1)}$ and $x^A_{(2)}$, respectively. Then, the sum over 5D null paths between 1 and 2, denoted here by ${\mathcal R}(x^A_{(1)}, x^A_{(2)})$, is positively defined, conformally invariant, and has the status of a microcanonical sum, determining the particle propagation between 1 and 2 \cite{Breban:2005a}. Null path integrals satisfy a selfconsistency relation virtue to their geometric interpretation
\begin{eqnarray}
\label{eq:self}
{\mathcal R}(x^A_{(1)}, x^A_{(2)})=\int\limits_{1\prec 3\prec 2}\mkern-18mu\; \sqrt{|h|}d^5x_{(3)} {\mathcal R}(x^A_{(1)}, x^A_{(3)}) {\mathcal R}(x^A_{(3)}, x^A_{(2)}).
\end{eqnarray} 

We postulate that observers perceive geometrically only four dimensions; the fifth dimension (e.g., $x^5$, $x^0$, etc.) manifests indirectly through its consequences. Thus, a complete description of the physics in the 5D space-time requires measurements of additional entities than 4D space-time events.  The interpretation of the 5D space-time by a 4D observer is greatly facilitated by symmetry. In particular, if the 5D geometry has a space-like Killing vector then the 5D physics can be interpreted as a 4D quantum mechanics, while if the 5D geometry has a time-like Killing vector then the 5D geometry can be interpreted as a 4D statistical mechanics \cite{Breban:2005a}. 

Starting from Eq.~\eqref{eq:self}, we gave a heuristic derivation for how 5D null path integrals satisfy the following partial differential equation \cite{Breban:2005a} 
\begin{eqnarray}
\partial^A\partial_A{\mathcal R}\equiv-\partial_0^2{\mathcal R}+\nabla^2{\mathcal R}+\partial_5^2{\mathcal R}=0,
\label{eq:F-KG}
\end{eqnarray}
in a flat 5D space-time with the metric $\eta_{AB}=\mbox{\it diag}(-1,1,1,1,1)$ in pseudocartesian coordinates $x^A$; $\partial_A$ stood for the partial derivative with respect to $x^A_{(2)}$.\footnote{It can also be shown that ${\mathcal R}$ satisfies Eq.~\eqref{eq:F-KG} in the first argument, $x^A_{(1)}$.} The KG equation \eqref{eq:KG} resulted from Fourier transforming Eq.~\eqref{eq:F-KG} with respect to $x^5$ and conjugating $x^5$ with the inverse Compton wavelength $\lambda^{-1}=mc/\hbar$.
Obtaining equations similar to \eqref{eq:F-KG} for path integrals in arbitrary curved space-times is cumbersome. 

Here we extend the discussion to a special class of 5D space-times that can be foliated into conformally flat 4D space-times (i.e., the 4D conformal factor is the inverse square lapse of the foliation) with superimposed electromagnetic fields, fit for describing many experimental setups. These space-times represent just local approximations of more realistic geometric constructs including non-trivial gravitational fields, whose metrics satisfy suitable field equations \cite{Wesson:2007ta,Breban:2005a}.  Hence, one should be cautious of anomalies that may result from certain choices of electromagnetic fields.

Removing the conformal factor from the metric as it is irrelevant for null-path counting, we have
\begin{eqnarray}
\label{eq:fol1}
\tilde h_{AB}=\left(\begin{array}{cc}
\eta_{\mu \nu}+\frac{q^2}{c^4} A_\mu A_\nu & \frac{q}{c^2} A_\mu \\
              \frac{q}{c^2} A_\nu   & 1
\end{array}\right),\quad \tilde h^{AB}=\left(\begin{array}{cc}
\eta^{\mu \nu}& -\frac{q}{c^2} A^\mu \\
              -\frac{q}{c^2} A^\nu   & 1+\frac{q^2}{c^4} A^\mu A^\nu 
\end{array}\right),
\end{eqnarray}
where the $\mu, \nu, ...$ indices go over $0,1,2,3$ and are raised with $\eta^{\mu\nu}$.  This space-time is obtained from the 5D flat one through a non-holonomic transformation of coordinates
\begin{eqnarray}
\label{eq:nonhol-noncyl}
dy^\mu&=&dx^\mu, \\
dy^5&=&dx^5-\frac{q}{c^2}A_\mu(x^\nu)dx^\mu.\nonumber
\end{eqnarray}
If $A_\mu dx^\mu$ is integrable, then the coordinate transformation is holonomic; i.e., the space-time remains flat, containing an electromagnetic field that is pure gauge. In the Lorentz gauge, applying transformation \eqref{eq:nonhol-noncyl} to Eq.~\eqref{eq:F-KG} yields
\begin{eqnarray}
\label{eq:KG-em-}
\left(\partial^\mu-\frac{q}{c^2}A^\mu\partial^5\right)\left(\partial_\mu-\frac{q}{c^2}A_\mu\partial_5\right){\mathcal R} +\partial^5\partial_5{\mathcal R} =0.
\end{eqnarray}
We further note that
\begin{eqnarray}
\label{eq:KG-em}
\left(\partial^\mu-\frac{q}{c^2}A^\mu\partial^5\right)\left(\partial_\mu-\frac{q}{c^2}A_\mu\partial_5\right){\mathcal R} +\partial^5\partial_5{\mathcal R}=\tilde\bigtriangledown^A\tilde\bigtriangledown_A{\mathcal R},
\end{eqnarray}
where $\tilde\bigtriangledown_A$ is the covariant derivative corresponding to the metric $\tilde h_{AB}$ (see Appendix \ref{ap:2}).
Hence, the 5D KG equation can be summarized in covariant fashion using 5D parallel transport alone
\begin{eqnarray}
\label{eq:KG-em--}
\tilde\bigtriangledown^A\tilde\bigtriangledown_A{\mathcal R} =0.
\end{eqnarray}

A 5D covariant probabilistic interpretation of Eq.~\eqref{eq:KG-em--} for the propagation of a single quantum particle remains challenging. However, a scalar probability density can be naturally defined. First, we note that the 5D null path intregrals ${\mathcal R}(x^A_{(1)}, x^A_{(2)})$ are positively defined for all pairs of events 1 and 2, $1\prec 2$. Second, Eq.~\eqref{eq:self} is formally identical to the selfconsistency relation of a conditional probability. Hence, up to proper normalization, we may intrepret ${\mathcal R}(x^A_{(1)}, x^A_{(2)})$ as the conditional probability that a particle at event 1 reaches event 2. The probability density that the particle is at event 2 is thus given by
\begin{eqnarray}
\frac{\left[\int\limits_{J_-(2)}\sqrt{|h|}d^5x_{(1)}{\mathcal R}(x^A_{(1)}, x^A_{(2)})\right]\left[\int\limits_{J_+(2)}\sqrt{|h|}d^5x_{(3)}{\mathcal R}(x^A_{(2)}, x^A_{(3)})\right]}{\iint\limits_{\forall 1\prec 3}|h|d^5x_{(1)}d^5x_{(3)}{\mathcal R}(x^A_{(1)}, x^A_{(3)})},
\end{eqnarray}
where $J_{\pm}(2)$ denote the causal future and past of event 2, respectively.  However, it is impossible to construct a probability 5-current density having this scalar field as its zeroth component.

\subsection{The KG equation in quantum mechanics}
\label{sec:qm}

The quantum mechanics picture of the 5D propagation applies in the case where the 5D metric is $x^5$-independent. Computing the path integral ${\mathcal R}(x^A_{(1)}, x^A_{(2)})$ in the 5D Lorentzian manifold with the metric $\tilde h_{AB}$ is now equivalent to computing the following integral 
\begin{eqnarray}
\Psi_{\pm}(\lambda^{-1};x^\mu_{(1)}, x^\mu_{(2)})=\int[d^4x]\exp\left[i\lambda^{-1}\int_{x^\mu_{(1)}}^{x^\mu_{(2)}}\left(\pm\sqrt{-\eta_{\mu \nu}dx^\mu dx^\nu}-\frac{q}{c^2}A_\rho dx^\rho\right)\right]
\end{eqnarray}
of paths between $x^\mu_{(1)}$ and $x^\mu_{(2)}$, the 4D projections of $x^A_{(1)}$ and $x^A_{(2)}$, respectively, in a 4D non-Lorentzian curved manifold with infinitesimal distance 
\begin{eqnarray}
ds_{4\;\pm}=\pm\sqrt{-\eta_{\mu \nu}dx^\mu dx^\nu}-\frac{q}{c^2}A_\rho dx^\rho,\nonumber
\end{eqnarray}
if the two neighboring events are causally ordered and $-ds_{4\;\pm}$ if they are inverse causally ordered \cite{Breban:2005a}. Hence, computing the path integral ${\mathcal R}(x^A_{(2)}, x^A_{(1)})$ is equivalent to computing $\Psi^*_{\pm}(\lambda^{-1};x^\mu_{(1)}, x^\mu_{(2)})$ \cite{Breban:2005a}, where $*$ is symbol for complex conjugation.  It is important to note that $\Psi_{\pm}$ is not a scalar field on the 4D manifold since a transformation of coordinates that reverses causality implies a complex conjugation of $\Psi_{\pm}$.

A differential equation for $\Psi_{\pm}(\lambda^{-1};x^\mu_{(1)}, x^\mu_{(2)})$ is obtained by Fourier transforming Eq.~\eqref{eq:KG-em-}, which results in the 4D KG equation for a particle in electromagnetic field \eqref{eq:KG-EM}. The probability density of localizing the quantum particle at a 4D space-time event may be constructed starting from the sum over loops in the 4D manifold
\begin{eqnarray}
{\mathcal P}(\lambda^{-1};x^\mu_{(1)}, x^\mu_{(2)})\equiv\Psi_{\pm}(\lambda^{-1};x^\mu_{(1)}, x^\mu_{(2)})\Psi_{\pm}(\lambda^{-1};x^\mu_{(2)}, x^\mu_{(1)})=\Psi_{\pm}(\lambda^{-1};x^\mu_{(1)}, x^\mu_{(2)})\Psi^*_{\pm}(\lambda^{-1};x^\mu_{(1)}, x^\mu_{(2)}).
\end{eqnarray}
Note that ${\mathcal P}$ is invariant under gauge transformations of the electromagnetic field and, just like ${\mathcal R}$, satisfies the self-consistency relation of conditional probability
\begin{eqnarray}
\label{eq:selfP}
{\mathcal P}(\lambda^{-1};x^\mu_{(1)}, x^\mu_{(2)})=\int\limits_{1\prec 3\prec 2}\mkern-18mu\; \sqrt{|\eta|}d^5x_{(3)} {\mathcal P}(\lambda^{-1};x^\mu_{(1)}, x^\mu_{(3)}) {\mathcal P}(\lambda^{-1};x^\mu_{(3)}, x^\mu_{(2)}),
\end{eqnarray} 
virtue to its geometrical interpretation. Hence, the probability density that a particle undergoing causal propagation is localized at the 4D event $x^\mu_{(2)}$ may be defined as
\begin{eqnarray}
\frac{\int\limits_{\forall 1\prec 2}\sqrt{|\eta|}d^4x_{(1)} \Psi_{\pm}(\lambda^{-1};x^\mu_{(1)}, x^\mu_{(2)})\Psi_{\pm}^*(\lambda^{-1};x^\mu_{(1)}, x^\mu_{(2)})}{\iint\limits_{\forall 1\prec 2}|\eta|d^4x_{(1)}d^4x_{(2)} \Psi_{\pm}(\lambda^{-1};x^\mu_{(1)}, x^\mu_{(2)})\Psi_{\pm}^*(\lambda^{-1};x^\mu_{(1)}, x^\mu_{(2)})},
\end{eqnarray}
and the probability density that a particle undergoing anti-causal propagation is localized at the 4D event $x^\mu_{(1)}$ may be defined as
\begin{eqnarray}
\frac{\int\limits_{\forall 1\prec 2}\sqrt{|\eta|}d^4x_{(2)} \Psi_{\pm}(\lambda^{-1};x^\mu_{(1)}, x^\mu_{(2)})\Psi_{\pm}^*(\lambda^{-1};x^\mu_{(1)}, x^\mu_{(2)})}{\iint\limits_{\forall 1\prec 2}|\eta|d^4x_{(1)}d^4x_{(2)} \Psi_{\pm}(\lambda^{-1};x^\mu_{(1)}, x^\mu_{(2)})\Psi_{\pm}^*(\lambda^{-1};x^\mu_{(1)}, x^\mu_{(2)})}.
\end{eqnarray}
However, like in Sec.~\ref{sec:5DKG}, a covariant form of the current density remains elusive.  In what follows, we explain the relationship between the KG and Schr\"odinger equations and discuss the implications for the probabilistic interpretation.

\subsubsection{Probabilistic interpretation for the KG equation without electromagnetic field}
Let us start with the KG equation for a free particle with 5-momentum $p_A$ (n.b., Eq.~\eqref{eq:F-KG} yields $p_Ap^A=0$) in a 5D flat space-time; i.e., Eq.~\eqref{eq:F-KG}. Using the light-cone coordinates 
\begin{eqnarray}
y^0=x^5-x^0\equiv c\tau,\quad y^j=x^j,\quad y^5=(x^0+x^5)/2,
\label{eq:change}
\end{eqnarray}
we obtain
\begin{eqnarray}
\frac{\partial}{\partial\tau}\frac{\partial}{\partial y^5}{\mathcal R} =-\frac{c}{2}\nabla^2{\mathcal R} .
\label{eq:a-S}
\end{eqnarray}
Performing a Fourier transform with respect to $y^5$ brings the KG equation given by Eq.~\eqref{eq:a-S} in the form of the Schr\"odinger equation
\begin{eqnarray}
i\frac{\partial}{\partial\tau}\hat\Psi=-\frac{c\hat\lambda}{2}\nabla^2\hat\Psi,
\label{eq:KGnewF}
\end{eqnarray}
where $\hat\Psi$ is the Fourier transform of ${\mathcal R} $ with respect to $y^5$ and $\hat\lambda^{-1}$ is conjugated to $y^5$; n.b., $\hat\lambda$ is defined over the whole real axis.  It is now straightforward to define a four-current density for the KG equation
\begin{eqnarray}
\label{eq:NJ-def}
\hat j_\tau=\hat\Psi^*\hat\Psi, \quad \hat j_k=\frac{c\hat\lambda}{2 i}(\hat\Psi^*\partial_k\hat\Psi-\hat\Psi\partial_k\hat\Psi^*),
\end{eqnarray}
satisfying
\begin{eqnarray}
\label{eq:NJ}
\frac{\partial\hat j^\tau}{\partial\tau}+\partial_k\hat j^k=0.
\end{eqnarray}
It is important to note that $(\hat j_\tau, \hat j_k)$ is {\it not} a four vector. While 5D and 4D covariance are lost in Eqs.~\eqref{eq:NJ-def} and \eqref{eq:NJ}, the probability density, $\hat j_\tau=\Psi^*\Psi$, has an appealing geometrical interpretation, in the spirit of the scalar probability density field discussed in Sec.~\ref{sec:qm}. 

The non-relativistic limit of Eq.~\eqref{eq:KGnewF} in the quantum mechanics picture, where an observer perceives $x^\mu$ geometrically and the momentum along $x^5$ as mass, is as follows. Taking the limit $|p_j|\ll |p_0|$ and $|p_j|\ll |p_5|$, we obtain $p_0\approx \pm p_5$ or $(\hbar/i)\partial_0\Psi \approx \pm p_5\Psi$; i.e., a non-relativistic particle starting at the origin of the coordinate frame is remains localized around $x^0\approx \pm x^5$ (see also the interpretation in Ref.~\cite{Breban:2005a}, Sec.~3.2). Thus, in the non-relativistic limit, $y^5\approx x^5$, $\lambda\approx mc/\hbar$ and Eq.~\eqref{eq:KGnewF} becomes the Schr\"odinger equation for a free particle
\begin{eqnarray}
i\hbar\frac{\partial}{\partial\tau}\Psi=-\frac{\hbar^2}{2m}\nabla^2\Psi,
\label{eq:S}
\end{eqnarray}
where $\tau$ is the coordinate time corresponding to the non-relativistic kinetic energy only.\footnote{The phase of a single wave of the Schr\"odinger equation is $-i[\tau p_jp^j/(2m)-p_jx^j]/\hbar$. The non-relativistic limit of the 5D phase yields $ip_Ax^A/\hbar\approx -i[t(E-mc^2)-p_jx^j]/\hbar\approx -i[tp_jp^j/(2m)-p_jx^j]/\hbar$; thus, the substitution $t\rightarrow \tau$ yields the non-relativistic phase.}  Equations \eqref{eq:NJ-def} and \eqref{eq:NJ} yield the well-known formulae for the probability and current density of the Schr\"odinger equation.  The probabilistic interpretation carries over naturally from the KG equation to the Schr\"odinger equation and yields the expected formalism.

\subsubsection{Probabilistic interpretation for the KG equation with electromagnetic field}
Equation \eqref{eq:KG-em-} can also be brought in a form that is linear in the derivative with respect to the zeroth coordinate and could serve for a 5D probabilistic interpretation. Straightforward algebra shows that this requires a coordinate transformation $x^A\rightarrow y^A$ such that $\partial y^0/\partial x^A$ is a 5D null vector; n.b., this condition is satisfied by Eqs.~\eqref{eq:change}. However, the reduction to 4D by a Fourier transform is generally not justified because $\tilde h_{AB}$ depends on $y^5$.

We restrict our discussion to the non-relativistic limit in the presence of weak electromagnetic fields. The coordinates $y^A$ \eqref{eq:change} are only approximate light-cone coordinates, but they suffice in this case. We obtain
\begin{eqnarray}
-\left(\partial_0-\frac{q}{c^2}A_0\partial_5\right)^2+\partial_5^2+\left(\partial_j-\frac{q}{c^2}A_j\partial_5\right)^2&=&-2\frac{\partial^2}{\partial y^0\partial y^5}+2\frac{q}{c^2}A_0\partial_0\partial_5-\left(\frac{q}{c^2}\right)^2A_0^2\partial_5^2\nonumber\\
&~&+\partial_j^2-2\frac{q}{c^2}A_j\partial_j\partial_5+\left(\frac{q}{c^2}\right)^2A_j^2\partial_5^2-\frac{q}{c^2}(\partial_\mu A^\mu)\partial_5.
\label{eq:qmchange}
\end{eqnarray}
Replacing the Lorentz gauge (i.e., $\partial_\mu A^\mu=0$) with the Coulomb gauge (i.e., $\partial_j A^j=0$), using $y^5\approx x^5$, and taking a Fourier transform with respect to $x^5$, Eq.~\eqref{eq:KG-em-} becomes
\begin{eqnarray}
\label{eq:Schrod}
-\frac{\hbar}{i}\frac{\partial \Psi}{\partial \tau}=\frac{1}{2m}\left(\frac{\hbar}{i}\nabla-\frac{mq}{c}\overrightarrow{A}\right)^2\Psi-\frac{i\hbar(mq)}{mc}A_0\partial_0\Psi+\frac{(qm)^2}{2mc^2}A_0^2\Psi.
\end{eqnarray}
In the limit of weak fields, we drop the term quadratic in $A_0$ and use the approximation $(\hbar/i)\partial_0\Psi \approx p_5\Psi$ in the term linear in $A_0$, which gives $-i\hbar(mq)A_0\partial_0\Psi/mc\approx (mq)A_0\Psi$. Hence, Eq.~\eqref{eq:Schrod} represents the Schr\"odinger equation, obtained as the non-relativistic limit of the KG equation. 

\subsection{The KG equation in statistical mechanics}
\label{sec:smetal}
We first analyze the case of the equation without electromagnetic field. A Laplace transform of Eq.~\eqref{eq:a-S} with respect to $\tau$ yields
\begin{eqnarray}
\frac{\partial}{\partial y^5}\psi=\frac{\Lambda}{2}\nabla^2\psi.
\label{eq:pFP}
\end{eqnarray}
With the notation $y^5\equiv cu$, in the non-relativistic limit where $y^5\approx x^5$, Eq.~\eqref{eq:pFP} becomes the Fokker-Planck equation (see \cite{Breban:2005a}, Sec.~4.2), the core of statistical mechanics
\begin{eqnarray}
\frac{\partial}{\partial u}\psi=\frac{c\Lambda}{2}\nabla^2\psi,
\label{eq:FP}
\end{eqnarray}
where $\Lambda=2/(\beta\zeta c)$, $\beta=1/(k_B T)$ and $\zeta$ is the drag coefficient.

We now discuss the non-relativistic limit in the presence of electromagnetic field. Statistical mechanics results in the particular case where the metric $\tilde h_{AB}$ is $x^0$-independent (see Ref.~\cite{Breban:2005a}, Sec.~4); thus, we further request that $\partial_0 A_\mu=0$.
Since the change of coordinates \eqref{eq:change} yields
\begin{eqnarray}
-\left(\partial_0-\frac{q}{c^2}A_0\partial_5\right)^2+\partial_5^2=-2\frac{\partial^2}{\partial y^0\partial y^5}+2\frac{q}{c^2}A_0\partial_0\partial_5-\left(\frac{q}{c^2}\right)^2[A_0^2\partial_5^2-A_0(\partial_5A_0)\partial_5]\nonumber,
\end{eqnarray}
assuming that the particle has a well-defined mass (i.e., $\partial_5A_\mu=0$), and the limit of weak fields bring the non-relativistic limit of Eq.~\eqref{eq:KG-em} in the following form\footnote{We make use of the zeroth order non-relativistic approximation $\partial_5\psi \approx\partial_0\psi $ in the terms that are first order in the electromagnetic field $A_\mu$.}
\begin{eqnarray}
2\frac{\partial^2}{\partial (cu)\partial y^0}{\mathcal R} =\left(\nabla-\frac{q}{c^2}\overrightarrow A\partial_0\right)^2{\mathcal R}+2\frac{q}{c^2}A_0\partial_0^2{\mathcal R}.
\label{eq:pFPem}
\end{eqnarray}
A Laplace transform with respect to time yields
\begin{eqnarray}
-\hbar\frac{\partial \psi}{\partial u}=\frac{-1}{2M}
\left(-\hbar\nabla+\frac{Mq}{c}\overrightarrow{A}\right)^2\psi-(Mq)A_0\psi,
\label{eq:pFPem_}
\end{eqnarray}
where $M$ is defined by $\Lambda\equiv\hbar/(Mc)$.

\subsection{Other interpretations of the KG equation}
Other changes of coordinates will bring the KG equation in the form of Eq.~\eqref{eq:a-S}, as well.  For example, the coordinate transformation
\begin{eqnarray}
y^0=x^3-x^0\equiv c\tau_z,\quad y^1=x^1,\quad y^2=x^2,\quad y^3=(x^0+x^3)/2,\quad y^5=x^5,
\label{eq:changez}
\end{eqnarray}
and, assuming that all fields are independent of $y^3$, a Fourier transform with respect to $y^3\equiv z$ yield, similarly to the calculations in Sec.~\ref{sec:qm}, a Schr\"odinger-like equation for ultrarelativistic particles.  We perform another  Fourier transform with respect to $x^5$ to make the equation easier to interpret
\begin{eqnarray}
i\hbar\left[1-\frac{2(mq)A_{z3}}{cp_z}\right]\frac{\partial}{\partial\tau_z}\Psi_z&=&\frac{c}{2p_z}\left[\left(\frac{\hbar}{i}\partial_1-\frac{mq}{c}A_{z1}\right)^2+\left(\frac{\hbar}{i}\partial_2-\frac{mq}{c}A_{z2}\right)^2\right]\Psi_z
\nonumber\\ &~&
+(mq)A_{z0}\Psi_z-\frac{(mq)^2}{cp_z}A_{z0}A_{z3}\Psi_z-\frac{m^2c^3}{2p_z}\Psi_z,
\label{eq:Sz}
\end{eqnarray}
where $A_{z\mu}$ is the transformed electromagnetic field
\begin{eqnarray}
A_{z0}=A_3-A_0\quad A_{z1}=A_1,\quad A_{z2}=A_2,\quad A_{z3}=(A_0+A_3)/2,
\label{eq:changeAz}
\end{eqnarray}
$p_z$ is the projection of the 5-momentum of the ultrarelativistic particle along the $z$-axis, and $\Psi_z$ is the Fourier transform of ${\mathcal R}$ with respect to both $z$ and $x^5$.  Equation \eqref{eq:Sz} describes the quantum motion of a particle of mass $m$ transverse to the $z$ direction and generalizes Eq.~(7.2) in Ref.~\cite{Breban:2005a}.

The current density has the following components
\begin{eqnarray}
\label{eq:NJ-defz}
j_{z\tau}&=&\Psi_z^*\Psi_z\left[1-\frac{2(mq)A_{z3}}{cp_z}\right]-\frac{2(mq)}{cp_z}\int d\tau_z\Psi_z^*\Psi_z\frac{\partial A_{z3}}{\partial\tau_z},\\ 
j_{z1}&=&\frac{c\hbar}{2 ip_z}(\Psi^*\partial_1\Psi-\Psi\partial_1\Psi^*),\\
j_{z2}&=&\frac{c\hbar}{2 ip_z}(\Psi^*\partial_2\Psi-\Psi\partial_1\Psi^*), 
\end{eqnarray}
and satisfies the 3D continuity equation.

\subsection{The hydrogenic atom from a 5D perspective}
The hydrogenic system is particularly suited for discussion here due to its extensively developed quantum theory in both the relativistic and non-relativistic regimes. The electron propagation problem can be set up in a 5D manifold with the metric $\tilde h_{AB}$ \eqref{eq:fol1} where the electromagnetic potential is $A_\mu=(Ze/x^jx_j,0_j)$. Hence, the following key physical quantities are introduced: the speed of light in vacuum $c$, the electron specific charge $q$, and, due to the translational symmetries of $\tilde h_{AB}$ along $x^5$ and $x^0$, two conjugate wavelengths $\lambda$ and $\lambda\sp{\prime}$. As mentioned previously, $\lambda$ stands for the Compton wavelength of the quantum particle; $\lambda\sp{\prime}\equiv\hbar c/E$ is associated to energy conservation.

\subsubsection{Energy spectrum}
The energy levels of the hydrogenic atom described by the KG equation are given by \cite{IZ}
\begin{eqnarray}
E_{nl}=mc^2\left\{1+\frac{Z^2\alpha^2}{[n-l-1/2+\sqrt{(l+1/2)^2+Z^2\alpha^2}]^2}\right\}^{-1/2},
\label{eq:HS-KG}
\end{eqnarray}
where $\alpha=e^2/(\hbar c)$ is the fine-structure constant, $n=1, 2,...$ and $l=0, 1,..., n$. The non-relativistic limit of Eq.~\eqref{eq:HS-KG} can be obtained formally as $\alpha\rightarrow 0$ \cite{IZ}. It is important to note that the principles of the KG equation do not provide a physical interpretation of the hydrogenic energy spectrum. The interpretation requires the principles of photonics, particularly the Ritz principle.

Using 5D physical quantities, Eq.~\eqref{eq:HS-KG} becomes
\begin{eqnarray}
\left[\left(\frac{\lambda\sp{\prime}_{nl}}{\lambda}\right)^2-1\right]\left[n-l-\frac{1}{2}+\sqrt{\left(l+\frac{1}{2}\right)^2+\left(\frac{\lambda^*}{\lambda}\right)^2}\;\right]^2=\left(\frac{\lambda^*}{\lambda}\right)^2,
\label{eq:HS-KG_}
\end{eqnarray}
where we introduced a new length scale, $\lambda^*\equiv q(Ze)/c^2$, resulting from the 5D metric parameterization and equal to the classical electron radius multiplied by $Z$; n.b., $\lambda^*/\lambda=Z\alpha$.  Thus, energy quantization \eqref{eq:HS-KG_} may be regarded as a {\it matching condition} between the metric length scale, $\lambda^*$, and the two wavelengths of particle propagation, $\lambda$ and $\lambda\sp{\prime}$.

In statistical mechanics, the coordinate $x^0$ is handled with a Laplace transform. To recover the traditional formalism of statistical mechanics, another Laplace transform is performed with respect to $x^5$ \cite{Breban:2005a}. The quantization condition becomes
\begin{eqnarray}
\label{eq:stat_q}
\left[\left(\frac{\Lambda}{\Lambda\sp{\prime}_{nl}}\right)^2-1\right]\left[n-l-\frac{1}{2}+\sqrt{\left(l+\frac{1}{2}\right)^2-\left(\frac{\lambda^*}{\Lambda\sp{\prime}_{nl}}\right)^2}\;\right]^2=-\left(\frac{\lambda^*}{\Lambda\sp{\prime}_{nl}}\right)^2.
\end{eqnarray}
Not only for the proof of concept, but also because it works well for the hydrogen atom, we take the non-relativistic limit of  Eq.~\eqref{eq:stat_q} 
\begin{eqnarray}
\Lambda\sp{\prime}_{nl}\approx\Lambda\left[1+\frac{1}{2n^2}\left(\frac{\lambda^*}{\Lambda}\right)^2\right],
\end{eqnarray}
which, with the notation $e_{nl}=\hbar c/\Lambda_{nl}$ and substituting $\Lambda$ and $\lambda^*$, becomes
\begin{eqnarray}
e_{nl}\approx Mc^2-\frac{Mc^2(Ze)^2q^2M^2}{2n^2\hbar^2c^2}\equiv e_n,
\end{eqnarray}
reminiscent of the traditional formula for the discrete hydrogenic energy levels.\footnote{In terms of temperature, the non-relativistic condition, $\lambda^*/\Lambda\ll 1$, becomes $T\gg (Ze)q\zeta/(2ck_B)$.}

\subsubsection{Canonical sum}
The energy levels obtained in the statistical interpretation may be used for constructing a canonical sum, defined as the trace of the Fokker-Planck propagator, integrated over all space-time (see Ref.~\cite{Breban:2005a}, Sec.~4.2). Its non-relativistic approximation can be obtained by adapting the calculations performed by Blinder for the traditional model of the hydrogenic atom \cite{1995JMP}.  

Integrating the Fokker-Planck propagator over all space-time leads to a divergent result. To maintain the dependence on volume in the canonical sum and, simultaneously, resolve the divergence problem, Blinder \cite{1995JMP} considered the hydrogen atom at the center of a large sphere of radius $R$ and performed the integration over the volume of the sphere.  He also assumed that, with good approximation, the energy eigenfunctions of the hydrogenic atom in infinite volume can be used for those of the atom trapped in the spherical cavity. Although appealing, this ansatz has important consequences for the interpretation of our statistical mechanics defined on a 4D Riemannian background. First, Blinder's ansatz may only be justified for the non-relativistic approximation since it breaks general covariance. Second, it requires a particular treatment of the canonical sum over continuous states. Our statistical mechanics describes particles at rest (i.e., whose initial and final events coincide). A free particle in an infinite volume is at rest if and only if it has vanishing kinetic energy. The situation is substantially different if the free particle is trapped in a cavity.  In this case, there exist at-rest particle states of non-vanishing kinetic energy: i.e., standing waves created within the cavity. However, computing the canonical sum over these states is outside the principled path described for our 4D covariant statistical mechanics; for this, we adopt Blinder's approach \cite{1995JMP}, as explained below.

We consider the canonical sum for the hydrogenic atom, $\mathcal{Z}$, split into the contributions due to the continuous and discrete parts of the energy spectrum, $\mathcal{Z}_{c,d}$, respectively 
\begin{eqnarray}
\mathcal{Z}=\mathcal{Z}_c+\mathcal{Z}_d.
\end{eqnarray}
Introducing 3D spherical coordinates, $\mathcal{Z}_c$ can be written as
\begin{eqnarray}
\mathcal{Z}_c=\int_0^R\! dr\,r^2 \int_0^\pi\! d\theta\,\sin(\theta)  \int_0^{2\pi}\!d\phi \int_0^\infty\! dk\,\sum_{l,m} |\psi_{klm}(r,\theta,\phi)|^2e^{-ue_k/\hbar},
\end{eqnarray}
where $e_k=Mc^2+\hbar^2k^2/(2M)$ are the eigenvalues of the continuous spectrum and $\psi_{klm}(r,\theta,\phi)$ are the corresponding eigenfunctions.
Similarly, $\mathcal{Z}_d$ is given by
\begin{eqnarray}
\mathcal{Z}_d=\int_0^R\! dr\,r^2 \int_0^\pi\! d\theta\,\sin(\theta)  \int_0^{2\pi}\!d\phi \sum_{n,l,m}|\psi_{nlm}(r,\theta,\phi)|^2e^{-ue_n/\hbar},
\end{eqnarray}
where $e_n$ are the eigenvalues of the discrete spectrum and $\psi_{nlm}(r,\theta,\phi)$ are the corresponding eigenfunctions. 

Adapting Blinder's results \cite{1995JMP} to our case, we obtain
\begin{eqnarray}
\mathcal{Z}_c=\frac{Ve^{-\eta_0}}{\Lambda^3(2\pi \eta_0)^{3/2}}-\frac{e^{-\eta_0}}{2}\sum_{n\geqslant 1}n^2\exp\left({\!-\frac{n^2\pi^{4/3}uc\Lambda}{2V^{2/3}}}\right)\left\{\exp\left[{\left(\frac{\lambda^*}{\Lambda}\right)^2\!\frac{\eta_0}{2n^2}}\right]\mbox{erfc}\left[\frac{\lambda^*}{\Lambda}\sqrt{\frac{\eta_0}{2n^2}}\,\right]-1\right\},
\label{eq:Zc}
\end{eqnarray}
where $\mbox{erfc}(\cdot)$ is the complementary error function, $\eta_0\equiv uMc^2/\hbar$ and $V$ is the volume of the sphere with radius $R$. Furthermore, we have
\begin{eqnarray}
\mathcal{Z}_d=\sum_{n\geqslant 1}e^{-ue_n/\hbar} \int_0^R\! dr D_n(r),
\label{eq:Zd}
\end{eqnarray}
where the function $D_n(\cdot)$ is given in terms of Whittaker functions $M_{n,1/2}(\cdot)$ and their derivatives with respect to the argument, denoted by primes,
\begin{eqnarray}
D_n(r)=\frac{4r^2}{n^3\rho^3}\left\{  \left[M'_{n,1/2}\left(\frac{2r}{n\rho}\right)\right]^2- M_{n,1/2}\left(\frac{2r}{n\rho}\right)M''_{n,1/2}\left(\frac{2r}{n\rho}\right)\right\}; \quad \rho\equiv\frac{\Lambda^2}{\lambda^*}.
\end{eqnarray}
In Appendix \ref{sec:conv_can}, we show that the series in Eqs.~\eqref{eq:Zc} and \eqref{eq:Zd} converge; our analysis of the function $D_n(\cdot)$ differs from Blinder's \cite{1995JMP}.

A thermodynamical interpretation of the hydrogenic atom may be proposed in analogy to that of a black-body cavity containing a single photon. One may think of the bound electron trapped in the hydrogenic potential as being macroscopically at rest, undergoing transitions in the rest mass, according to a discrete spectrum. 

To complete the interpretation of the 4D statistical picture, we need to provide a formula for the quantum of physical time $u$ in terms of experimentally accessible quantities. This is done empirically; in Ref.~\cite{Breban:2005a}, a comparison to Feynman's version of statistical mechanics in the case of free particles led to $u=2m/\zeta$.  While this result may be adopted here in the case where $Ze=0$, it is unclear whether it holds in the case of the hydrogenic atom. Comparison between the thermodynamical results resulting from $\mathcal{Z}$ and experiment may offer an estimate of $u$, even though spin-related effects are prominent in experiments. This remains a topic for future work. 

\section{Discussion and conclusion}

In 1934, Pauli and Weisskopf \cite{Pauli:1934dy} provided a physical interpretation for the KG equation that is still in use today.  Within the formalism of quantum field theory in Minkovski space-time, they interpreted $j_\mu$ \eqref{eq:j} as current density of electric charge.  They used complex conjugation to transform particle states into antiparticle states and vice versa (i.e., according to the Feynman-Stueckelberg interpretation whereby an antiparticle is s particle propagating backward in time). They further showed that $j_\mu$ consists of two terms, one due to particles and another one due to antiparticles, such that a complex conjugation of the states changes $j_\mu$ into $-j_\mu$.

The interpretation of the KG equation proposed here has common features to that proposed earlier by Pauli and Weisskopf \cite{Pauli:1934dy}.  Particularly, we used that causal and anti-causal propagators are linked by complex conjugation.  This feature appeared naturally in our theory, owing to a 4D non-Lorentzian space-time.  This space-time occurred from foliating a 5D Lorentzian manifold, while imposing that the count of null paths between two points, 1 and 2, in the 5D Lorentzian manifold is linked by Fourier transformed to a sum over time-like paths between the 4D projections of the points 1 and 2 in the non-Lorentzian space-time \cite{Breban:2005a}.

In our 4D non-Lorentzian space-time, a transformation of coordinates that reverses causality (i.e., turns particles into antiparticles according to the Feynman-Stueckelberg interpretation) is equivalent to a complex conjugation of the KG wavefunction and, therefore, a change in sign of $j_\mu$.  In this regard, it may be justified to interpret $j_\mu$ as current density of electric charge.  

Following previous work, we discussed physical interpretations resulting from breaking covariance of a 5D wave equation, which is satisfied by the count of null paths in certain 5D Lorentzian space-times. When the 5D metric is independent of the fifth, space-like coordinate, the KG equation resulted naturally by foliation and Fourier transform, as previously noted for the flat space-time \cite{Seahra:2001vi}.  To take the non-relativistic limit of the KG equation in a flat space-time, we used light-cone coordinates to bring the 5D wave equation in the form of the Schr\"odinger equation.  This approach has two advantages: (a) transparent non-relativistic limit, and (b) natural (yet, non-covariant) probabilistic interpretation of the KG equation.  We further showed how the Schr\"odinger equation resulted as the non-relativistic limit of the KG equation in the case of weak electromagnetic fields.  Then we discussed other interpretations of the 5D wave equation in the case where the 5D metric is not only independent of $x^5$, but also $x^0$ and $x^3$, respectively.  As an application of this formalism, we discussed the spectrum of the hydrogenic atom.  Using the fact that the 5D wave equation can be interpreted as a Focker-Planck equation in the case where the 5D metric is independent of both $x^5$ and $x^0$, we calculated the canonical sum of the hydrogenic atom.

\appendix

\section{A 5D covariant form of the KG equation in certain space-times}
\label{ap:2}
We assume that the null path integral ${\mathcal R}(x^B)$ satisfies the following equation defined on the surface of a 5D null cone 
\begin{eqnarray}
\label{eq:KG-gen}
\tilde\bigtriangledown^A\tilde\bigtriangledown_A{\mathcal R} (x^B)=0,
\end{eqnarray}
where $\tilde\bigtriangledown_A$ is the covariant derivative of the manifold with the metric $\tilde h_{AB}$.
We show that Eq.~\eqref{eq:KG-gen} yields the well-known KG equation in the presence of electromagnetic field \eqref{eq:KG-EM}. Considering the path integral ${\mathcal R}(x^B)$ to be scalar with respect to the covariant derivative, we have
\begin{eqnarray}
\label{eq:KG-op}
\tilde\bigtriangledown^A\tilde\bigtriangledown_A{\mathcal R} (x^D)=\tilde h^{AB}\partial_A\partial_B{\mathcal R}(x^D)+\tilde h^{AB} \tilde\Gamma^C_{AB}\partial_C{\mathcal R} (x^D), 
\end{eqnarray}
where $\tilde\Gamma^C_{AB}$ are the Christoffel symbols of the metric $\tilde h_{AB}$; see \cite{Wesson:2007ta}, Sec.~5.6 for the general formulae. Using the notation $N_\mu\equiv-\frac{q}{c^2}A_\mu$, the right term in the RHS can be further expanded as
\begin{eqnarray}
\tilde h^{AB} \tilde\Gamma^C_{AB}\partial_C= \tilde h^{\mu\nu}\tilde\Gamma_{\mu\nu}^C\partial_C+2\tilde h^{\mu 5}\tilde\Gamma^C_{\mu 5}\partial_C=\eta^{\mu\nu}\tilde\Gamma_{\mu\nu}^\rho\partial_\rho+\eta^{\mu\nu}\tilde\Gamma_{\mu\nu}^5\partial_5+2N^\mu\tilde\Gamma_{\mu5}^\rho\partial_\rho+2N^\mu\tilde\Gamma_{\mu5}^5\partial_5,\nonumber
\end{eqnarray}
where
\begin{eqnarray}
\arraycolsep=6pt\def\arraystretch{1.5}
\begin{array}{ll} \eta^{\mu\nu}\tilde\Gamma_{\mu\nu}^\rho\partial_\rho=N^\mu(\partial_\mu N^\nu-\partial^\nu N_\mu)\partial_\nu, &
\eta^{\mu\nu}\tilde\Gamma_{\mu\nu}^5\partial_5=-(\partial_\mu N^\mu)\partial_5,\\
2N^\mu\tilde\Gamma_{\mu5}^\rho\partial_\rho =-\eta^{\mu\nu}\tilde\Gamma_{\mu\nu}^\rho\partial_\rho, &
2N^\mu\tilde\Gamma_{\mu5}^5\partial_5=0.
\end{array}\nonumber
\end{eqnarray}
Equation \eqref{eq:KG-op} becomes 
\begin{eqnarray}
\label{eq:KG-}
\tilde\bigtriangledown^A\tilde\bigtriangledown_A{\mathcal R}=[\partial^\mu\partial_\mu+2N^\mu\partial_\mu\partial_5+(N_\mu N^\mu+1)\partial_5\partial^5-(\partial_\mu N^\mu)\partial_5]{\mathcal R}=0.
\end{eqnarray}
Applying a Fourier transform with respect to $x^5$ and re-grouping the differential operators, we obtain
\begin{eqnarray}
\label{eq:KG-final}
\left(\partial^\mu-i\frac{qm}{c\hbar}A^\mu\right)\left(\partial_\mu-i\frac{qm}{c\hbar}A_\mu\right)\Psi-\frac{m^2c^2}{\hbar^2}\Psi-2i\frac{qm}{c\hbar}(\partial_\mu A^\mu)\Psi=0,
\end{eqnarray}
where $\Psi$ is the Fourier transform of ${\mathcal R}$ with respect to $x^5$, and $m$ is the mass of the quantum particle.  In the Lorentz gauge, Eq.~\eqref{eq:KG-final} represents the KG equation for a particle with electric charge $qm\equiv e$ in electromagnetic field \eqref{eq:KG-EM}.  The above derivation makes use only of the covariant derivative of a 5D manifold (rather than gauge covariant derivatives) where the electromagnetic and gravitational field are on equal footage.  In fact, the electromagnetic field appears as a gauge field only when the 5D manifold is given a 4D interpretation.

\section{Convergence of the canonical sum}
\label{sec:conv_can}
\begin{figure}[h]
\begin{centering}
\includegraphics[width=.5\textwidth]{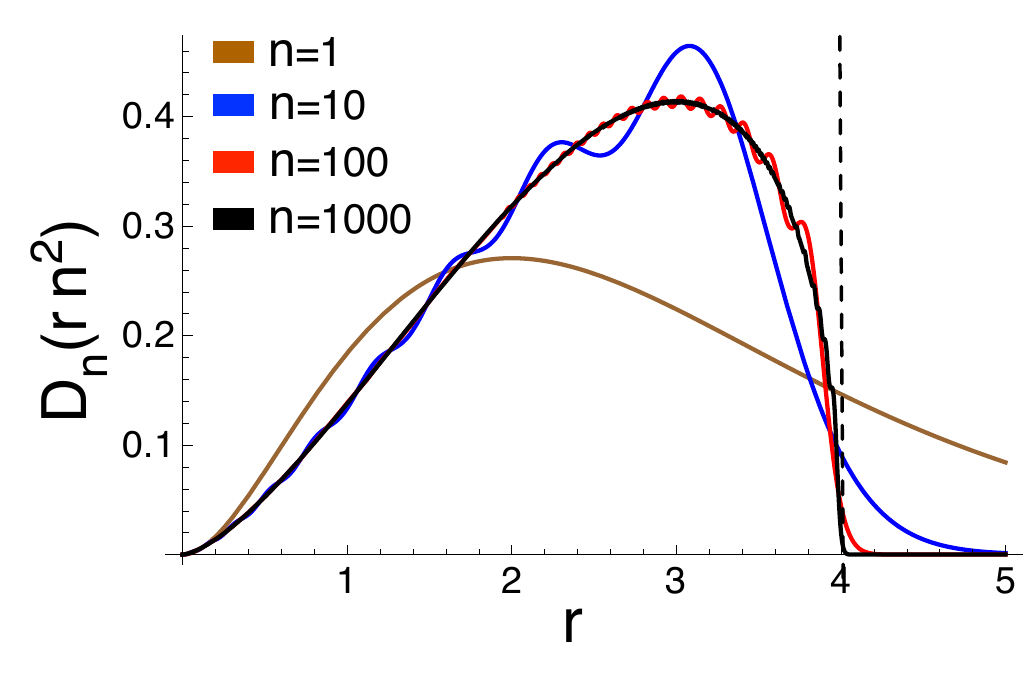}
\caption{\label{fig:1} $D_n(rn^2)$ versus $r$ for $n=1$, 10, 100 and 1000.}
\end{centering}
\end{figure}

$\mathcal{Z}_c$ is obviously convergent since the factor within braces in Eq.~\eqref{eq:Zc} approaches $-[\lambda^*/(n\Lambda)]\sqrt{2\eta_0/\pi}$ as $n\rightarrow \infty$. An argument for the convergence of $\mathcal{Z}_d$ is as follows. The function $D_n(\cdot)$ is normalized according to 
\begin{eqnarray}
\int_0^\infty \!dr\,D_n(r)=n^2,
\end{eqnarray}
being interpreted as the degeneracy of the $n$th discrete energy level \cite{Blinder:1993ud}.  Hence, we immediately obtain 
\begin{eqnarray}
\int_0^\infty \!dr\,D_n(rn^2)=1.
\end{eqnarray}
In fact, Blinder noted that, as $n\rightarrow \infty$, $D_n(r)$ approaches a universal reduced form \cite{Blinder:1993ud}. In particular, we have
\begin{eqnarray}
D_n(rn^2)\xrightarrow{n\rightarrow\infty}D(r).
\end{eqnarray}
Figure \ref{fig:1} illustrates graphs of $D_n(rn^2)$ versus $r$ for $n=1$, 10, 100 and 1000, suggesting our next result.\vspace*{\baselineskip}

{\bf Proposition.} Consider $\rho/2$ for the unit length. The universal function $D:[0,\infty)\rightarrow[0,\infty)$ is given by
\begin{eqnarray}
\label{eq:Duniv}
D(r)=\left\{\begin{array}{l}r^{3/2}\sqrt{4-r}/(4\pi)\\ 0\end{array}\begin{array}{l}\mbox{if }r<4;\\\mbox{if } r\geqslant 4.\end{array} \right. 
\end{eqnarray}

{\bf Proof.} We proceed by taking the limit of 
\begin{eqnarray}
\label{eq:Dn}
D_n(rn^2)=\frac{r^2n}{2}\left[M'_{n,1/2}(rn)^2-M_{n,1/2}(rn)M''_{n,1/2}(rn)\right]
\end{eqnarray}
as $n\rightarrow \infty$. An asymptotic form for the Whittaker function $M_{n,1/2}(nr)$ can be obtained via Laguerre polynomials $L_n^{(1)}(nr)$ since
\begin{eqnarray}
\label{eq:Whitt-Lag}
M_{n,1/2}(rn)=re^{-rn/2}L_{n-1}^{(1)}(rn).
\end{eqnarray}
Asymptotic expansions for the Laguerre polynomials developed by Erd\'elyi \cite{Erdelyi:1960}, later extended by Muckenhoupt \cite{Muckenhoupt:1970uv}, followed by expansions for the resulting Bessel and Airy functions, yield
\begin{eqnarray}
\label{eq:Asymp}
L_{n-1}^{(1)}(rn)\sim\left\{\begin{array}{l}\frac{\displaystyle e^{rn/2}}{\displaystyle r\sqrt{\pi n\varrho'(r/4)}}\cos[4n\varrho(r/4)-3\pi/4]\\ \\\frac{\displaystyle (-1)^n e^{rn/2}}{\displaystyle r\sqrt{\pi n\varsigma'(r/4)}}e^{-4n \varsigma(r/4)} \end{array}\begin{array}{l}\mbox{if }r<4;\\ \\ \\ \mbox{if } r\geqslant 4.\end{array} \right.
\end{eqnarray}
where
\begin{eqnarray}
\label{eq:Asymp.extras}
\varrho(r)\equiv\frac{1}{2}\left(\sqrt{r-r^2}+\sin^{-1}\sqrt{r}\right),\quad
\varsigma(r)\equiv\frac{1}{2}\left(\sqrt{r^2-r}-\cosh^{-1}\sqrt{r}\right).
\end{eqnarray}
Plugging Eqs.~\eqref{eq:Asymp} and \eqref{eq:Asymp.extras} into Eq.~\eqref{eq:Whitt-Lag}, we obtain the required asymptotic form of $M_{n,1/2}(nr)$ for taking the limit of Eq.~\eqref{eq:Dn}. Straightforward calculation yields Eq.~\eqref{eq:Duniv}. $\square$

{\bf Remark.} It is also straightforward to see that $D(r)$ is normalized, $\int_0^\infty\! dr\, D(r)=\int_0^4\! dr\, D(r)=1$.

The convergence behavior of $\mathcal{Z}_d$ can be described using $D(r)$
\begin{eqnarray}
\mathcal{Z}_d\sim \mathcal{C}+\sum_{n\geqslant n_0}e^{-ue_n/\hbar}\; n^2\!\!\int_0^{R/n^2}\!\! dr\, D(r).
\end{eqnarray}
The effect of considering the hydrogenic atom in a large spherical cavity becomes now transparent. For $n\leqslant\sqrt{R/4}$, the degeneracy of the energy levels remains $\sim n^2$; otherwise, the degeneracy is gradually reduced through the incomplete integral over $D(r)$. In particular, we have
\begin{eqnarray}
n^2\!\!\int_0^{R/n^2}\!\! dr\, D(r)\xrightarrow{n\rightarrow\infty}\frac{R^{5/2}}{5\pi n^3}.
\end{eqnarray}
Thus, since $e_n$ approaches $Mc^2$ as $n\rightarrow \infty$, $\mathcal{Z}_d$ converges like a Dirichlet series.
\end{document}